\documentclass[pra,twocolumn,superscriptaddress]{revtex4-1}

\usepackage[utf8]{inputenc} 
\usepackage{graphicx}
\usepackage{dcolumn}
\usepackage{amssymb,amsmath}
\usepackage{bm}
\usepackage{xcolor}
\usepackage{float}

\usepackage{hyperref}
\hypersetup{
  colorlinks   = true, 
  urlcolor     = blue, 
  linkcolor    = blue, 
  citecolor   = blue 
}

\definecolor{bleuQ}{HTML}{009CB4}

\newcommand{ \parO }[1]{\left( #1 \right)}
\newcommand{ \parS }[1]{\left[ #1 \right]}

\newcommand{ \sy }[0]{\sigma_y}

\newcommand{ \Hm }[0]{\mathcal{H}}

\newcommand{ \bv }[0]{\mathbf{b}}
\newcommand{ \Bv }[0]{\mathbf{B}}

\newcommand{ \sigv }[0]{\boldsymbol{\sigma}}

\newcommand{ \drm }[0]{\mathrm{d}}

\begin{document}


\title{Optimized micromagnet geometries for Majorana zero modes in low g-factor materials}

\author{Sara Turcotte}
\email{sara.turcotte@usherbrooke.ca}

\author{Samuel Boutin}
\author{Julien Camirand Lemyre}
\author{Ion Garate}
\affiliation{
 Institut quantique et D\'epartement de physique, Universit\'e de Sherbrooke, Sherbrooke, Qu\'ebec J1K 2R1, Canada
}

\author{Michel Pioro-Ladri\`ere}%
\affiliation{
 Institut quantique et D\'epartement de physique, Universit\'e de Sherbrooke, Sherbrooke, Qu\'ebec J1K 2R1, Canada
}
\date{\today}

\begin{abstract}
Solid-state experimental realizations of Majorana bound states are based on materials with strong intrinsic spin-orbit interactions.
In this work, we explore an alternative approach where spin-orbit coupling is induced artificially through a non-uniform magnetic field that originates from an array of micromagnets. 
Using a recently developed optimization algorithm, 
 we find suitable magnet geometries for the emergence of topological superconductivity in wires without intrinsic spin-orbit coupling.  
We confirm the robustness of Majorana bound states against disorder and periodic potentials whose amplitudes do not exceed the Zeeman energy.
Furthermore, we identify low g-factor materials commonly used in mesoscopic physics experiments as viable candidates for Majorana devices. 
\end{abstract}

\maketitle

\section{Introduction}
Topologically protected states, such as Majorana zero modes, are envisioned as building blocks for hardware efficient quantum computation~\cite{Nayak:2008ty,Sarma:2015kk}. 
One of the most promising implementations relies on semiconducting nanowires with strong spin-orbit coupling and proximity-induced superconductivity~\cite{Oreg:2010xy,Lutchyn:2010rm,Lutchyn:2017qy}.
In this setup, a magnetic field can drive a topological phase transition, where Majorana bound states (MBS) emerge as localized states at the ends of the nanowire~\cite{Aguado2017a}.
Recent experiments have shown increasingly convincing signatures of these modes in InAs~\cite{Albrecht:2016kq,Deng:2016wt} and InSb~\cite{Mourik:2012ee,Chen2017,Zhang2018a} nanowires.
But despite progress in growth techniques~\cite{Lutchyn:2017qy}, scaling the nanowire approach to the two-dimensional networks needed for quantum computation remains a daunting task ~\cite{Alicea:2011ad,Karzig2017,Plugge:2017dg}. 

Motivated by the aforementioned difficulty, new top-down fabrication methods for InAs/Al heterostructures are under development~\cite{Suominen2017,Shabani:2016qv,Vaitiekeinas2018}. 
Likewise, exploring other classes of planar materials could lead to new prospects for MBS-based devices. 
For example, the low disorder of GaAs/AlGaAs or Si/SiGe heterostructures could be beneficial for topological protection~\cite{Alicea:2012fj}, and the well-established fabrication capabilities of materials such as silicon would be an asset for the development of complex devices~\cite{Zwanenburg2013}. 
Even though induced superconductivity was demonstrated in several of these low g-factor materials~\cite{Chiodi2017,Wan:2015qv,Heersche2007}, the absence of a strong spin-orbit coupling hinders the possibility of a topologically protected phase.
This lack of intrinsic spin-orbit coupling can be remedied with the help of an inhomogeneous magnetic texture~\cite{Choy:2011wq}, engineered by the use of magnetic adatoms~\cite{Nadj-Perge:2013cr}, arrays of micromagnets~\cite{Kjaergaard:2012tg,Maurer2018}, magnetic domains~\cite{KontosGroup}, or magnetic tunnel junctions~\cite{Zhou2019,Fatin:2016gp,Matos-Abiague2017}  placed in proximity to the wire. 
The latter was also proposed as an alternative approach for the braiding of MBS.

In this work, we focus on architectures in which magnetic textures are created by micromagnet arrays.
Although the idea of using magnetic textures for the engineering of topological phases is widely recognized~\cite{Choy:2011wq,Nadj-Perge:2013cr,Kjaergaard:2012tg,Klinovaja2012b,Rainis:2014le}, relatively little effort has been devoted towards modeling realistic magnet arrays and understanding the effect of non-helical magnetic fields on MBS~\cite{Zhou2019,Maurer2018}.
In addition, two important obstacles of the micromagnet approach have been largely overlooked. 
First, the small g-factors in semiconductors such as Si, Ge or GaAs/AlGaAs heterostructures limit the Zeeman energy and therefore make it more difficult to reach the topological phase. 
Second, this approach is subjected to an additional source of disorder due to the possible misplacement of micromagnets during nanofabrication.

The main objective of the present paper is to demonstrate that robust topological superconductivity can be engineered in low g-factor materials placed in proximity to realistic micromagnet arrays, with the crucial help of an automated process~\cite{Boutin2018} that determines the optimal magnet arrangement.  
In Sec.~\ref{sec:model_and_optim}, we introduce a single-channel model describing the Majorana wire and review the method used to optimize the shape and position of the micromagnets. This simple model helps to develop a physical understanding of the outcome of the numerical optimization procedure. 
In Sec.~\ref{sec:arrayprop}, we investigate three simple magnet geometries and assess their robustness against variations of tunable parameters such as the chemical potential and the external magnetic field. We also highlight the detrimental impact of periodic electrostatic potentials generated in the nanowire by the micromagnets.
Section~\ref{sec:disorder} focuses on the impact of possible micromagnet misplacements arising during the fabrication process, and identifies geometries that are resilient to this additional disorder channel. 
In Sec.~\ref{sec:materials} we analyze the engineering of MBS in various low g-factor materials, and discuss the influence of micromagnets on wires with strong intrinsic spin-orbit coupling.
Finally, in Sec.~\ref{sec:discussion}, we investigate multi-band effects and confirm that they do not alter the main conclusions extracted from the single-band model.
\clearpage


\section{Model and methods}\label{sec:model_and_optim}

\subsection{Single-band nanowire model}\label{subsec:model}

We consider a single-band nanowire of length $L$ with proximity-induced s-wave superconductivity. 
This wire could be a genuine nanowire or else the result of electrostatic gating in a two-dimensional electron gas (2DEG) formed at the interface of a semiconductor heterostructure. 
Although below we will concentrate on the latter scenario, our optimization procedure can also be applied to micromagnets placed in the vicinity of semiconductor nanowires.

The low energy physics of the wire is described by the Hamiltonian 
\begin{equation}
	H = H_0 + H_\Delta,
	\label{eq:Htot}
\end{equation}
where
\begin{equation}
	H_0 = \int_0^L \drm x \, \psi^\dag(x) \Hm_0(x) \psi(x) 
	\label{eq:H0}
\end{equation}
is the non-superconducting part, $\psi^{(\dag)}(x)$ is a two-component spinor that annihilates (creates) an electron at position $x$, and
\begin{equation}
	H_{\Delta} = \int_0^L \drm x \,  \parS{\Delta \psi_\uparrow^\dag(x) \psi_\downarrow^\dag(x) 
	+ h.c.
	}
	\label{Hdelta}
\end{equation}
is the superconducting part with the proximity-induced s-wave gap $\Delta$. In Eq.~\eqref{eq:H0}, we have defined
\begin{equation}
	\Hm_0(x) = \frac{ p_x^2 }{2m^*} - \mu + \frac{ 1 }{2}g \mu_B \Bv(x)\cdot \sigv ,  
	\label{eq:Hm0}
\end{equation}
with $p_x=-i\hbar \partial_x$, $m^*$ the effective mass, $\mu$ the chemical potential, 
$g$ the effective $g$-factor, $\mu_B$ the Bohr magneton and ${\boldsymbol\sigma}$ a vector of spin Pauli matrices. 
In addition, the total local magnetic field reads 
\begin{equation}
	\Bv(x) = \Bv_0 + \bv(x),
\end{equation}
where we have separated an external uniform magnetic field $\Bv_0$ from the magnetic texture $\bv(x)$ that is generated by polarized micromagnets (see Sec. \ref{sec:arrayprop}). 
We simulate cobalt micromagnets with a magnetization of $M=1.8$~T parallel to a polarization field of amplitude $|\mathbf{B}_0|=0.2$~T~\cite{Lachance-Quirion:2015rw}.
The non-uniform magnetic field can be calculated either analytically (in the case of bar magnets~\cite{Engel-Herbert2005}) or else using finite elements calculations~\footnote{Using e.g. the RADIA \textit{Wolfram Mathematica} package \url{http://www.esrf.eu/Accelerators/Groups/InsertionDevices/Software/Radia}}.
Unless otherwise specified, we focus on silicon with $m^* = 0.2$ (in units of the bare electron mass) and $g=2$. 
For this material, we neglect the weak intrinsic spin-orbit interaction~\cite{Zwanenburg2013,Maurer2018} and consider a wire of length $L=5$~$\mu$m. 
The interplay between intrinsic spin-orbit coupling and magnetic textures will be considered in Sec.~\ref{sec:materials}.
In that case, Eq.~\eqref{eq:Hm0} must be modified to include the additional term $\alpha p_x\sy$, where $\alpha$ is the intrinsic spin-orbit coupling strength.

As shown in e.g. Ref.~\cite{Kjaergaard:2012tg}, a magnetic field rotating in space is unitarily equivalent to the action of a uniform magnetic field and spin-orbit coupling oriented perpendicular to the field rotation plane.
In the simple magnet geometries considered below, the field rotation will take place on a single plane.
Then, the amplitude of the artificial spin-orbit coupling can be written as
\begin{equation}
\alpha _\mathrm{eff} = \frac{\hbar}{2m^*} \frac{d \phi}{d x},
\label{eq:alpha_eff}
\end{equation} 
where $\phi$ is the magnetic field angle. 
In the case of a perfect spiral field of period $p$, with $\bv~=~b_0 \parS{\cos \parO{ 2\pi x /p } \hat{\mathbf{x}} + \sin \parO{ 2\pi x /p }\hat{\mathbf{z}}}$,
the artificial spin-orbit coupling is uniform along the wire and has an amplitude \mbox{$\alpha_{\mathrm{eff}} = \hbar \pi/mp$}. 

Depending on the material used to form the 2DEG, various techniques can be employed to induce superconductivity in the channel \cite{,Chiodi2017,Wan:2015qv,Heersche2007}. 

Superconductivity may be induced through two superconducting contacts at each extremities of the nanowire~\cite{KontosGroup}.
One challenge of this platform for transport measurement is that one of the superconducting leads must be replaced by a normal metallic lead in order to be able to measure the zero-bias peak in the tunneling conductance. 
Another challenge is that the Majorana modes in such a 
superconductor/normal/superconductor structure may overlap strongly even when the length of the normal part exceeds the superconducting coherence length~\footnote{We thank Yuval Oreg for bringing this point to our attention.}. 
In this so-called long junction regime, 
the induced energy gap at zero phase bias is $\sim\hbar v/L$ in the ballistic limit, where $v$ is the Fermi velocity and $L$ is the length of the normal segment. 
The characteristic localization length of a Majorana bound state is then $\sim L$, i.e. comparable to the length of the normal wire, leading to a strong hybridization of the two MBS at the extremities regardless of the length of the wire, unless a phase bias of $\pi$ is applied~\cite{Lopes2018}.
In order to avoid these issues, superconductivity may instead be induced laterally on the nanowire, along its entire length, following an approach similar to that of Ref. [\onlinecite{Wan:2015qv}].

Experimentally, the amplitude of the proximity-induced superconducting gap will depend on multiple parameters such as the semiconductor-superconductor interface transparency, the applied external magnetic field and the superconductor thickness~\cite{Tinkham:2004uq}. 
As these parameters are sensitive to experimental details, hereafter we adopt the conservative estimate of $\Delta = 16.5~\mu$eV $\approx 200$~mK, which is approximately half the superconducting gap reported for doped silicon at zero external field in Ref.~\cite{Chiodi2017}. 
This value of the induced superconducting gap is an order of magnitude smaller than the one commonly reported in one dimensional semiconducting nanowires coated with epitaxial aluminum.
For simplicity, we approximate $\Delta$ to be independent of the magnetic field texture.
This approximation is justified when the critical field of the bulk superconductor far exceeds $|{\bf B}(x)|$, a circumstance that will be realized when using niobium, for example.\\

In the following sections, we characterize the MBS by diagonalizing numerically a discretized version of 
Eq.~\eqref{eq:Htot}.
From the diagonalized Hamiltonian, we extract (i) the energy gap $\Delta_0$, also referred to as the topological gap, which separates the zero modes from the low-lying quasiparticle excitations, and (ii) the energy splitting $\epsilon_M$ between the two MBS. 
The latter quantity gives a quantitative measure of the MBS localization ($\epsilon_M$ decreases as the overlap between the MBS wave functions is reduced)~\cite{Alicea:2012fj}. 
Together, $\Delta_0$ and $\epsilon_M$ characterize the topological protection of the MBS and provide bounds on timescales for braiding operations in future Majorana-based qubits~\cite{Sarma:2015kk}.

 \begin{figure*}[t]
\centering
\includegraphics[width=1\textwidth]{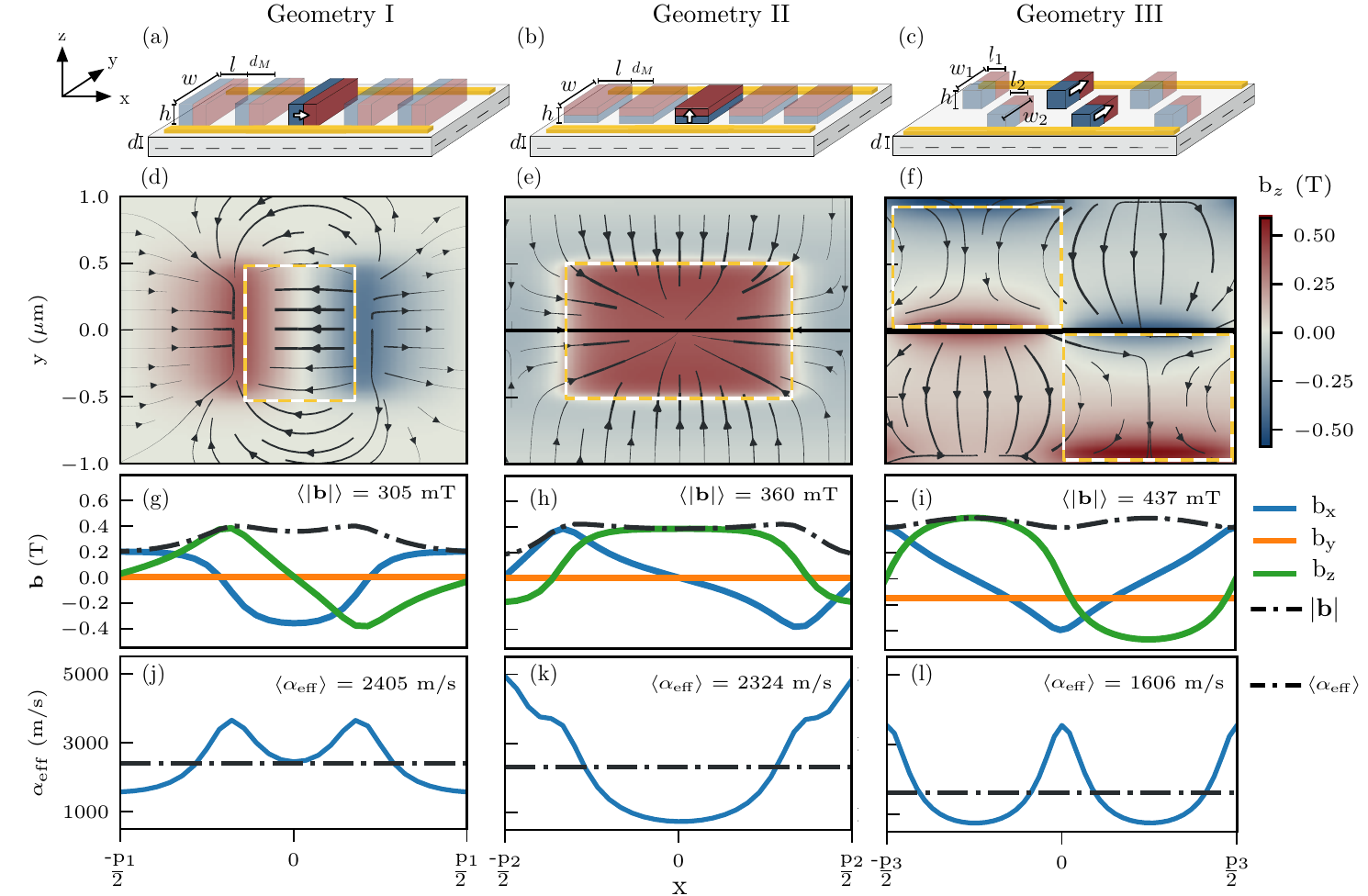}
\caption{\label{fig:3 geom} (a-c) Schematics of the three studied geometries with all micromagnets polarized along the same direction. The magnets form arrays of periodicity $p$. Electrostatic gates (yellow) are used to deplete the two-dimensional electron gas located at a depth $d$ below the heterostructure surface. A uniform external field is applied in the polarization direction. For clarity, the superconductor is not shown. (a) Magnets polarized along the $x$ axis, with parameters $l$ = $h$ = 265 nm, $w$ = 1 $\mu$m, $d_M$ = 490 nm, and $p$ = 755 nm. (b) Magnets polarized along $z$ axis, where $l$ = 520 nm, $w$ = 1 $\mu$m, $h$ = 280 nm, $d_M$ = 250 nm, and $p$ = 770 nm. (c) Magnets polarized along $y$, with $l_{1,2}$ = 575 nm, $w_1$ = 960 nm, $w_2$ = 930 nm, $h$ = 575 nm, $y_1$ = 480 nm, $y_2$ = -465 nm, and $p$ = 1.15 $\mu$m. (d-f) Magnetic field lines within a unit cell, at $d=50$ nm. The color map describes the $z$ component of the field. The black line at $y= 0$ represents the nanowire position, while the white and yellow dashed lines represent magnets in the unit cell. (g-i) Profiles of all three components of the magnetic field, at $d=50$ nm. The dashed line denotes the magnitude of {\bf b}.  
(j-l) The corresponding effective spin-orbit coupling generated by {\bf b$(x)$}.
The dashed line denotes the spatial average of $\alpha_{\mathrm{eff}}$ over a unit cell.
}
\end{figure*}

\subsection{Optimization method}\label{subsec:optim}

Finding the optimal spatial configuration of the micromagnets, which will lead to the largest $\Delta_0$ and smallest $\epsilon_M$, is a nontrivial task.
Here, we optimize micromagnet geometries following the RGF-GRAPE algorithm~\cite{Boutin2018}, which is based on an analogy between the recursive Green's function (RGF) method~\cite{Thouless:1981wq} used for quantum transport calculations and the gradient ascent pulse engineering (GRAPE) algorithm of quantum optimal control~\cite{Khaneja:2005tw}. RGF-GRAPE allows for an efficient gradient-based optimization of functions of local (on-site) retarded Green's functions. 
This optimization procedure uses a gradient descent to minimize (maximize) the bulk energy gap when the wire is in the trivial (topological) phase~\cite{Boutin2018}. As a result, irrespective of the initial values of the tunable parameters, the algorithm converges towards regions in parameter space that are deep in the topological regime. 


More concretely, the optimization procedure is carried out by minimizing the product of two quantities: (i)~the topological character of the wire and (ii)~a measure of the bulk gap.
The topological phase is characterized using the so-called \emph{topological visibility} $Q$~\cite{DasSarma:2016ad}.
For a finite-size superconducting wire breaking time-reversal symmetry (class D~\cite{Chiu:2016kq}), this quantity can be obtained from the scattering matrix as $Q = \det r$, where $r$ is the zero-energy reflection matrix in the Majorana basis. 
The $\mathbb{Z}_2$ topological invariant characterizing the phase is then simply $\mathcal{Q}=\mathrm{sign}(Q)$, with $\mathcal{Q} = \pm1$ in the trivial (+1) and topological (-1) phase~\cite{Akhmerov:2011pi}.
For numerical efficiency, instead of the bulk gap we optimize the inverse of the localization length of zero energy states near the extremities of the wire~\cite{Boutin2018}.
By considering a nanowire strongly coupled to metallic leads, 
zero-energy modes will leak in the wire on a length scale that 
is inversely proportional to the bulk gap of the wire \footnote{Note, however, that the optimization of the localization length is not always equivalent to the optimization of the energy gap. For instance, for a fixed energy gap, it is possible to reduce the localization length by clustering energy eigenvalues together.}.
This localization length is obtained from the zero-energy local density of states. 
As a side product, in the topological phase, the algorithm will also minimize the localization length of the MBS, which will generally lead to a reduction of the zero-mode splitting $\epsilon_M$.
Although the inclusion of metallic leads is necessary to this algorithm, the optimized parameter configurations are robust to changes of the boundary conditions.

We refer the reader to Ref.~\cite{Boutin2018}  for an extensive discussion concerning the details of the optimization algorithm. 
Here, we simply mention some differences in the implementation of the algorithm between Ref.~\cite{Boutin2018}  and the present work.
First, we use a basin hopping global optimization method~\cite{Wales:1997sf}. 
This algorithm implements a series of gradient-based optimizations separated by stochastic perturbations to the optimization solution, which allows to explore a larger portion of parameter space and reduce the risk of finding low quality local extrema.
Second, we perform the optimization simultaneously for wires with different chemical potentials and require that all of them attain the topological phase. 
This favors solutions where the topological phase is stable on a larger chemical potential range, a desirable outcome for experimental implementations.
Finally, since we consider a relatively small optimization parameter space (see Sec. \ref{sec:arrayprop}-\ref{sec:materials}), we employ a simpler finite difference gradient calculation instead of the analytical gradient used in Ref.~\cite{Boutin2018}. 


\section{Micromagnet arrays }
\label{sec:arrayprop}

\subsection{Optimized geometries} \label{subsec:arrayopt}
In the following, we focus on geometries that meet realistic experimental constraints. In particular, we consider geometries where all magnets are polarized in the direction of the external magnetic field $\Bv_0$.
This design choice leads to rotating magnetic field textures while circumventing the need for complex magnet arrays that would contain either materials with different magnetization profiles or small single-domain magnets arranged in an anti-parallel fashion~\cite{Schuh2015}.
For the latter configuration, an external field exceeding the coercive field would align all magnets and potentially ruin the field texture needed to attain MBS.
Such constraint is absent in the geometries we consider, thereby enabling a larger parameter space for the engineering of MBS.

The three magnet arrangements (I, II and III) we study are depicted in Fig.~\ref{fig:3 geom} ({a-c}).
In all three geometries, an array of micromagnets and electrostatic gates are placed at a distance $d$ above a 2DEG. 
The gates produce the desired confinement potential to form a single band nanowire in the 2DEG.  
The magnets are polarized along the $x$-, $z$-, and $y$-axes for geometries I, II and III (respectively).
We note that geometry III presents the added advantage of having the external field aligned along the easy axis of the magnets. 
The lower polarization field of the magnets in this configuration extends the tuning range of the external field.

\begin{figure}[t]
	\centering
	\includegraphics[width=1.0\linewidth]{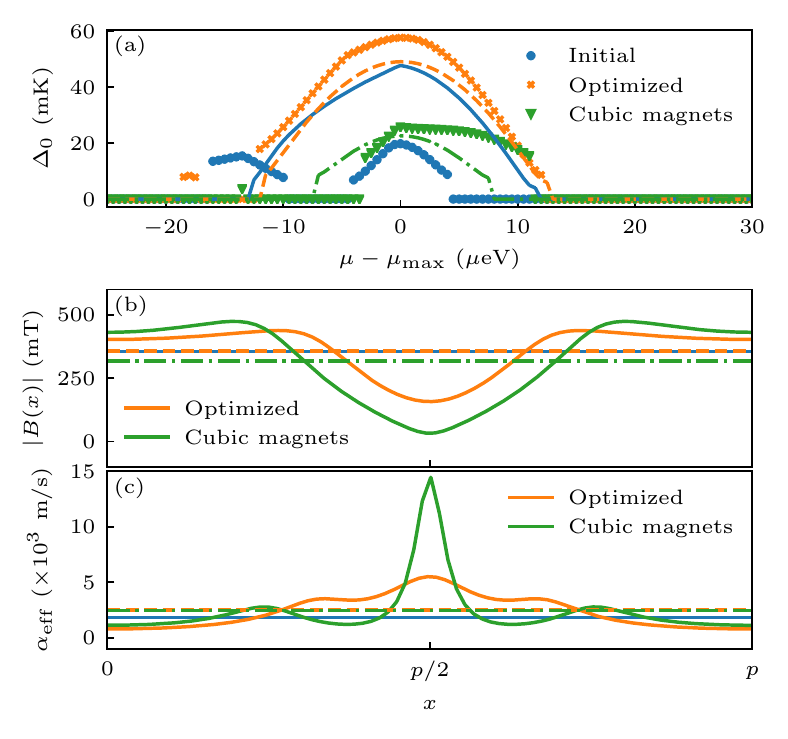}
	\caption{Topological gap, total magnetic field amplitude and effective spin-orbit coupling for geometry I.
	        Blue, orange and green curves correspond respectively to the starting point of the optimization, to the outcome of the optimization and to an array of cubic magnets with the same periodicity as the optimized geometry.
		(a)
		Topological gap as a function of the wire chemical potential, with $\Delta_0 = 0$ indicating the trivial phase ($\mathcal{Q}>0$). 
The chemical potential is offset so that the maximal value of $\Delta_0$ is at $\mu-\mu_{max}= 0~\mu$eV. 
The circles, crosses and triangles are the results for magnet arrays. The lines are the results for a perfectly spiral magnetic texture, whose amplitude and periodicity correspond to the average value of the magnetic field and spin-orbit coupling in the magnet array.
		(b) Amplitude of the total magnetic field for a unit cell of {length $p=$~755~nm}.
		(c)
		Effective spin-orbit coupling strength calculated from Eq.~\eqref{eq:alpha_eff}.
		The orange dashed (green dotted-dashed) horizontal lines indicate the average of the curves over a unit cell, while the 
blue lines are the average values for the initial array. Even though the average Zeeman and spin-orbit energies vary little in the course of the optimization, the gain in the topological gap is significant.
		%
	}
	\label{fig:optim_example}
\end{figure}

An automated optimization of each geometry allows us to systematically take into account constraints on the design of the magnet arrays. 
Thus, we restrict the optimization to experimentally realizable identical magnets with dimensions\mbox{ $h$, $w$, $l$,\mbox{ $d_M$}} between 50 nm and 1 $\mu$m and a conservative aspect ratio of \mbox{$h/\max(l,w)\leq 1$} (see Fig.~\ref{fig:3 geom} for parameter definitions){ \cite{Ye1995,Skuras2001}. The later constraint leads to a competition between minimizing the period of the array (maximizing the effective spin-orbit coupling) and maximizing the amplitude of the magnetic texture by increasing the height of the magnets. On the contrary, the other bounds on the magnet dimensions have a negligible impact on our results.
Although non-periodic arrays might give rise to improved topological properties \cite{Boutin2018}, we restrict the optimization to periodic arrays containing one or two micromagnets per unit cell. 
Thus, the starting point for the optimization is a periodic array of cubic magnets with dimensions \mbox{$h, w, l, d_M=500$~nm}.
Aperiodicities originating from fabrication errors will nevertheless be accounted for in Sec.~\ref{sec:disorder}.

Figures~\ref{fig:3 geom} {(d-i)} display the results of the optimization procedure for each geometry. 
An external magnetic field of $|\Bv_0|=200$~mT, necessary to polarize the magnets,  was included in the optimization. 
While this additional field can be detrimental to the emergence of Majorana modes~\cite{Maurer2018}, this is not the case if taken into account during the design of the magnet arrays. 

As an example, for geometries I and III, the optimization procedure naturally finds magnet arrays where the texture is offset in order to compensate for the external field. This offset lead to a helical magnetic texture of larger amplitude than in the case with $|\mathbf{B}_0 |=0$, thus increasing the topological gap. 
Geometry II appears to make the design of such an offset more difficult and no solution of this type is found by the optimization procedure.
The resulting spin-orbit coupling calculated from Eq.~\eqref{eq:alpha_eff} (without the polarizing field) is shown in Fig.~\ref{fig:3 geom} {(j-l)} and highlights the differences in spin-orbit profile for geometries with similar average Zeeman energies. The three geometries present an average spin-orbit coupling greater than 1600~m/s. While this is an order of magnitude lower than typical values observed in InAs or InSb nanowires, it does not constitute the limiting factor to reach the topological phase (see Sec.~\ref{subsec:Robustness}). 

To better appreciate the benefits of the optimization procedure, Fig.~\ref{fig:optim_example}~(a) compares the topological gap as a function of the wire chemical potential for the initial and optimized arrays in geometry I.
The optimization leads to an increase of the topological gap by a factor of \mbox{$\sim 3$} and the topological phase is reached for a significantly wider range of the chemical potential.
Even with optimized profiles, the topological gap is smaller than the one reported in semiconductor nanowires coated with epitaxial Al~\cite{Grivnin2019}. 
A larger value of $\Delta$ is unlikely to improve our topological gap significantly, mainly because the amplitude of ${\bf b}(x)$ is bounded by the saturation  magnetization of the micromagnets. It is nonetheless possible that more complex geometries of micromagnets will be conducive to larger topological gaps.   

As the optimized array has a smaller unit cell than the initial one, it is natural to wonder whether the observed improvement of the topological gap is solely the result of a larger effective spin-orbit coupling (a $35\%$ increase). 
To investigate this, we compare the optimized array to an array of cubic magnets with the same periodicity.
While the increased spin-orbit coupling due to the reduced period does lead to an increased topological gap, the array of cubic magnets falls short from the optimized array. 
As shown in Fig.~\ref{fig:optim_example}~(b), this can be understood by the 15\% larger Zeeman energy of the optimized array compared to the cubic magnets.
Another advantage of the optimized array with respect to the cubic array is that its magnetic texture partly compensates the external magnetic field. 
This compensation, beneficial for the topological gap, leads to smoother magnetic field profiles [see Fig.~\ref{fig:optim_example}~(b,c)].


Finally, Fig.~\ref{fig:optim_example}~(a) shows that the dependence of $\Delta_0$ on $\mu$ is similar in the optimized array and in a hypothetical wire with spatially uniform Zeeman and spin-orbit energies, provided that the latter are chosen to be equal to the spatially averaged values $\langle \alpha_{\mathrm{eff}}\rangle$ and $g \mu_B \langle |\textbf B|\rangle$ of the optimized array.
This behavior is reproduced in optimized arrays belonging to geometries II and III.

\subsection{Robustness over  parameter variations}
\label{subsec:Robustness}

It is important to assess whether or not the optimized geometries obtained in Sec.~\ref{subsec:arrayopt} exhibit robust MBS over modest changes of parameters such as the chemical potential, the external magnetic field and the induced superconducting gap.
Indeed, a geometry presenting a high sensitivity to such parameters would be difficult to implement experimentally, due to a limited precision in attaining the parameter values. 

Although the engineered magnetic textures are not perfectly helical, we can gain useful intuition by comparing numerical results to the analytic expressions for this ideal case. 
In a long nanowire without intrinsic spin-orbit coupling placed under a helical magnetic field and without any uniform external field, the condition for the appearance of MBS reads 
\begin{equation}
\frac12g\mu_B |{\mathbf{b}}| > \sqrt{|\Delta|^2 + (\mu -\tilde{\mu})^2},
\label{eq:topocond}
\end{equation}
where $\tilde{\mu}$ is a shift in the chemical potential due to the effective spin-orbit coupling and $|{\mathbf{b}}|$ the amplitude of the spiral field~\cite{Kjaergaard:2012tg}. 
From Eq.~\eqref{eq:topocond}, we anticipate that the topological phase will be realized in a larger interval of chemical potential when the amplitude of the Zeeman energy is increased. 
This expectation is confirmed by Fig.~\ref{fig:optimizationmu} (a), where geometry III displays the widest range of $\mu$ for which $\Delta_0>0$.
It is easy to understand why the range of $\mu$ for which the topological phase remains robust is at most a few tens of $\mu{\rm eV}$. 
The local magnetic fields in the nanowire are several hundreds of mT. Together with a g-factor of order unity, translate into Zeeman gaps of a few tens of $\mu{\rm eV}$ in the normal-state energy spectrum of the nanowire. In order to attain the spinless fermion regime necessary for the realization of the topological phase, the chemical potential must be placed within this Zeeman gap. This highlights the benefit of maximizing the Zeeman energy for a given array of micromagnets, through optimization \footnote{The Zeeman energies we find are larger than the ones predicted in Ref.~[\onlinecite{Maurer2018}] for related geometries.}.

\begin{figure}[tb]
\includegraphics[width=0.45\textwidth]{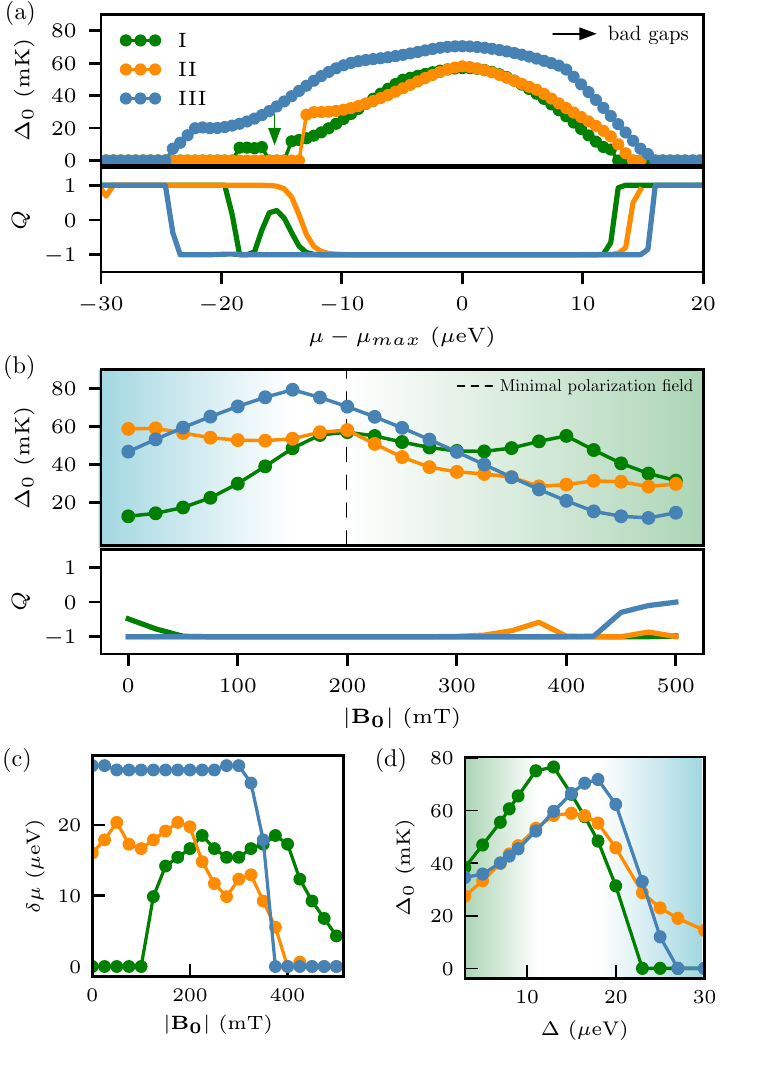}
\caption{\label{fig:optimizationmu} Robustness of the topological gap under the variation of parameters in all three geometries.  
(a) Topological gap $\Delta_0$ and topological visibility ($Q$)as function of the chemical potential at $B_0=$~200~mT. Here $\Delta_0=0$ represents the trivial phase, characterized by the the topological invariant ($\mathcal{Q}=+1$ ($Q>0$)). See section \ref{subsec:optim} for further details. Apparent discontinuities in the topological gap as function of the chemical potential are due to finite-size effects; they gradually disappear if the length of the wire is made increasingly longer than the period of the magnetic texture.
For each geometry, the plots are offset by $\mu_{\text{max}}$ for the sake of comparison. 
(b) Magnetic field and topological visibility ($Q$) dependence of $\Delta_0$ at $\mu-\mu_{max}=0$. 
(c) Dependence of $\delta\mu$ (namely the interval of chemical potential for which $\Delta_0 >30$~mK) on the external magnetic field.  
(d) Effect of the superconducting proximity gap $\Delta$ on the the topological gap. 
The shaded areas in panels (b) and (d) represent the regions where the topological gap is limited by $\Delta_1$ (blue) or $\Delta_2$ (green); see main text for definitions.}
\end{figure}

The intuition from the ideal helical case can be extended to understand the dependence of the topological gap on various parameters. 
In the ideal case, the topological protection is determined from the energy gaps at zero momentum ($k=0$) and at the Fermi momentum ($k=k_f$). 
Following the notation of Ref.~\cite{Cayao2017}, we define the gap at $k=0$ in the $\mu \approx \tilde{\mu} $ limit as
\begin{equation}
 	\Delta_1 = \frac{1}{2}g\mu_B|b-b_c|,
 	\label{d1}
\end{equation} 
 where $b_c = 2\Delta/(g \mu_B)$ is the critical field at which the topological phase transition occurs [cf. Eq.~\eqref{eq:topocond}]. 
The gap at $k=k_f$ is given by 
\begin{equation}
	\Delta _2 \approx   \frac{2\Delta}{[ 2+\sqrt{(g \mu_B b/2)^2/E_{\mathrm{so}}^2 +4} ]^{1/2}},
	\label{d2}
\end{equation}
 where $E_{\mathrm{so}} = m^*\alpha^2_{\mathrm{eff}}/2$ is the spin-orbit energy scale. 
The topological gap is dictated by the smallest between $\Delta_1$ and $\Delta_2$.
As we enter the topological phase from lower magnetic fields ($g \mu_B b \gtrsim \Delta$),  $\Delta_0$ is limited by $\Delta_1$. 
Deeper in the topological phase ($g \mu_B b\gg\Delta$), the topological gap becomes limited by $\Delta_2$, which decreases as the ratio $g \mu_B b/E_{\rm so}$ increases. 
The latter effect follows from a reduction of the effective $p$-wave superconducting gap, due to the alignement of spins at $\pm k_f$ as the Zeeman field is increased. 

The preceding observations are also relevant to wires placed in proximity to micromagnet arrays, as evidenced by Fig.~\ref{fig:optimizationmu} (b). 
For geometries I and III, the additional  Zeeman energy at low external field brings the wire deeper into the topological phase (i.e. $\Delta_0$ grows with $B_0$) by increasing $\Delta_1$. 
At higher $B_0$, the suppression of $\Delta_2$ brings about a decrease in $\Delta_0$.
For geometry II, the effective spin-orbit coupling is weak enough to have $\Delta_1 > \Delta_2$ at $B_0 = 0$ T. 
In this case,  $\Delta_0$ is limited by $\Delta_2$. 
Accordingly, $\Delta_0$ is quite insensitive to the external field at low $B_0$, and decreases as $B_0$ is made stronger. 
At high $B_0$, the function $\Delta_0(B_0)$ is nevertheless more complex than in uniform wires with perfect spiral fields, because the band structure is distorted by the non-uniform spin-orbit coupling and the appearance of undesired gaps. 



Suppressions and revivals of the topological gap are revealed in Fig.~\ref{fig:optimizationmu} (a) as the chemical potential is varied. 
Maurer \textit{et al.} showed~\cite{Maurer2018} that undesired gaps (so-called \emph{bad gaps}) appear in the band structure when the average Zeeman energy over a unit cell is nonzero. 
When the chemical potential enters a bad gap, the wire becomes a trivial insulator ($\Delta_0=0$).
For all three geometries, the polarizing field offsets one component of the total magnetic field, thus creating a nonzero average field in a unit cell. 
From there, increasing the amplitude of the external field augments the energy spans for undesired gaps. 
As expected from the discussion of Fig.~\ref{fig:3 geom}(g-i), this effect is most striking for geometry II where the texture cannot compensate the external magnetic field.
This is further highlighted in Fig.~\ref{fig:optimizationmu} (c), where we define $\delta\mu$ as the chemical potential range over which $\Delta_0>30$~mK. 
Sharp steps in $\delta\mu \, (B_0)$  are indicative of the opening of bad gaps in the band structure, and the fastest reduction of $\delta\mu$ is observed for geometry II. 
We also note that finite size effects contribute to the reduction of $\delta\mu$ at $B\gtrsim 300$~mT due to the overlap of the MBS wave functions.

In low g-factor materials, the interplay between $\Delta_0$ and $\Delta$ can be crucial for the observation of MBS (see  Fig.~\ref{fig:optimizationmu} (d)).  
All geometries show robust MBS for values of $\Delta$ that are close to those reported experimentally in low g-factor materials at zero external field \cite{Chiodi2017,Wan:2015qv}.
Far from the crossover between $\Delta_1$-limited and $\Delta_2$-limited regions, the topological gap varies roughly linearly with $\Delta$.
This agrees with the behavior that one would expect in a uniform wire. 
However, the non-helical character of the magnetic texture becomes evident in the fact that the slope in the $\Delta_1$-limited region (at higher $\Delta$) is not the same for all geometries.

\begin{figure}[tb]
  \includegraphics[width=0.45\textwidth]{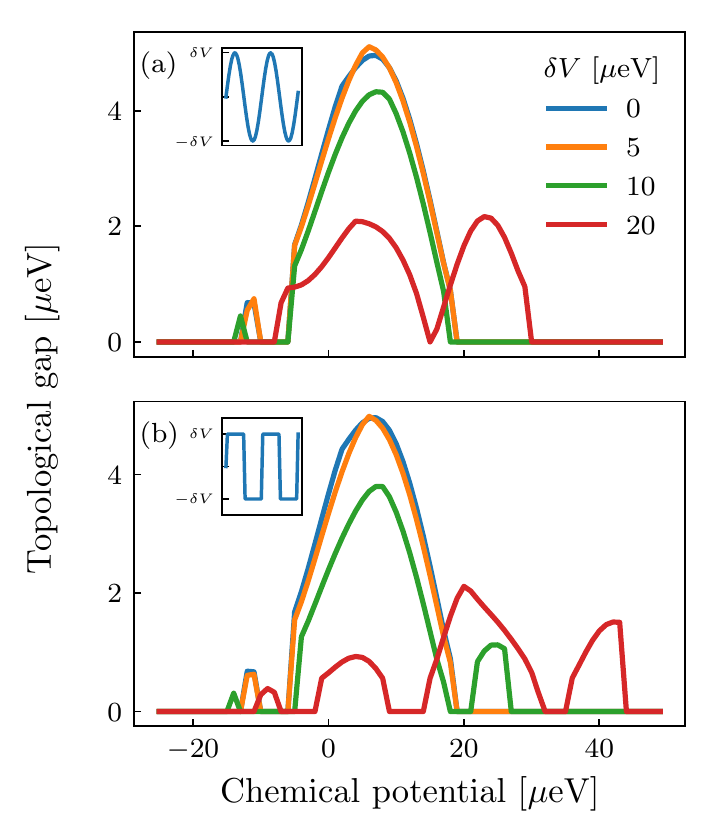}
  \caption{\label{fig:periodic_potential} Topological gap in the presence of
(a) a smooth potential modulation $\delta V \sin(2\pi x /p)$ and
(b) a sharp potential modulation $\delta V \,\text{sign}[\sin(2\pi x /p)]$, with $p$ the period of the magnetic texture and $\delta V$ the amplitude of the modulation. Calculations were performed using the parameters of Fig.~\ref{fig:optim_example} which uses geometry I.
}
  \end{figure}
%


\subsection{Electrostatic potential modulations}
The presence of micromagnets in a mesoscopic device can modify the electrostatic environment and lead to modulations of the electrostatic potential. 
While a rigorous electrostatic simulation is beyond the scope of this work, we have incorporated to our calculation a periodic electrostatic potential (with the same periodicity as the micromagnet array) in order to study its effect on the topological phase. 
We have found that the periodic potential leads to two main effects.
First, it opens “bad gaps” in the electronic structure, which reduce the interval of the chemical potential for which the spinless fermion regime can be realized.
Second, it flattens the electronic dispersion, because electrons tend to get localized in the troughs of the potential.
Both of these effects are harmful for the realization of Majorana modes. Figure~\ref{fig:periodic_potential} presents the effect of both a smooth and sharp potential modulation on the topological gap. We find that the Majorana modes and the topological gaps remain robust if the amplitude of the potential variations is comparable to or smaller than the Zeeman energy scale ($\sim 10 \ \mu{\rm eV}$). 
Moreover, the robustness is augmented if one optimizes the magnet geometry in the presence of a periodic potential.\\

Electrostatic modulations can have multiple origins. On the one hand, chemical potential shifts arising from variations of the work function or trap-states density could be partially compensated by the application of a uniform gate voltage on the magnet array. On the other hand, strain from lattice mismatch or different dilatation coefficient between the magnets and the semiconductor can induce large potential shifts \cite{Thorbeck2015}. While a realistic simulation of this effect is beyond the scope of this paper, it highlights the need to mitigate strain in device engineering. Improving  fabrication techniques or materials growth can be key to achieve low-strain magnet deposition. Another method is to physically separate the magnet array from the substrate to alleviate any strain at the interface. This could be achieved by changing the geometry to the one of a suspended nanowire  \cite{KontosGroup} or a two-dimensional electron gas \cite{Okazaki2016} hanging close to a magnet array.

 


\section{Majorana modes with disorder}
\label{sec:disorder}

In a real device, errors in micromagnet patterning give rise to disordered arrays with non-periodic magnetic field profiles. 
In this section, we investigate the robustness of MBS to such fabrication noise.
Specifically, we focus on the impact of random deviations from the optimized array designs 
on two key quantities: (i) the topological gap ($\Delta_0$), and (ii) the energy splitting between the MBS ($\epsilon_M$). 

We model the experimental variations by allowing deviations from the previously optimized dimensions and positions of each magnet in the array.
These deviations are sampled from a gaussian distribution with a standard deviation of 20~nm.

We assume the deposition of the cobalt magnets to be made in a single step. Since the deposition can be made uniform over the area containing the magnets (5 $\mu$m $\times$ 2 $\mu$m), all magnets have the same height. 
The chemical potential for a wire with a disordered array is fixed to the optimized value obtained for an array without noise. 

\begin{figure}[t]
\includegraphics[width=0.45\textwidth]{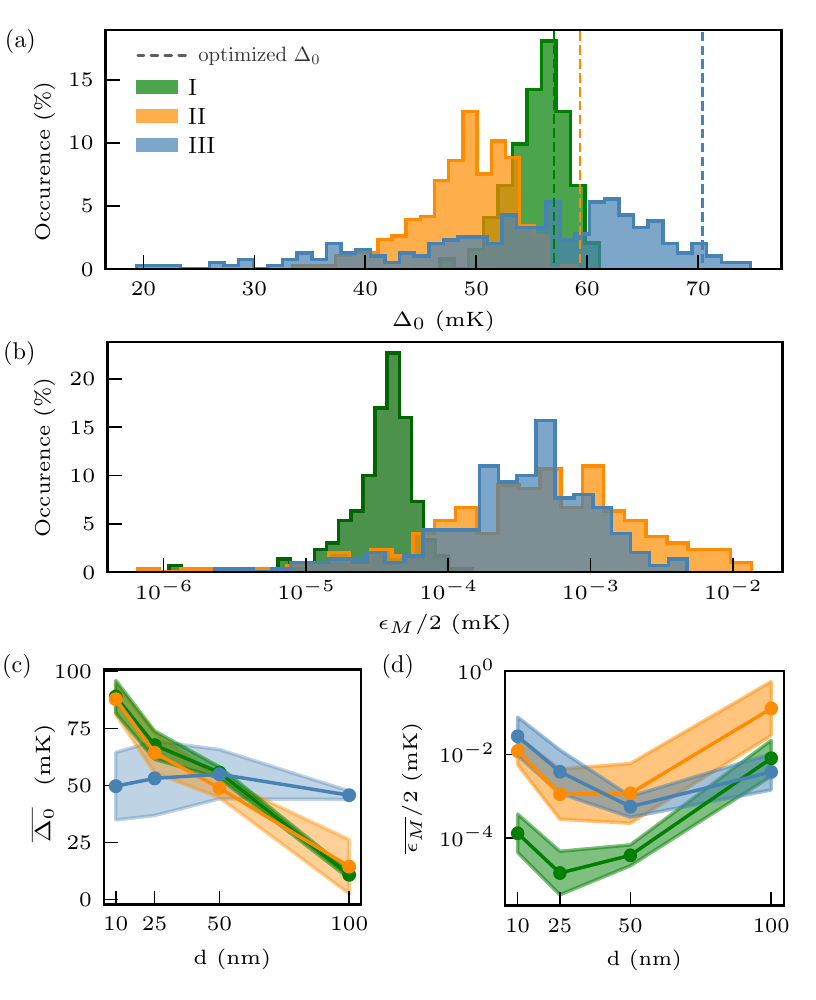}
\caption{\label{fig:optimization} 
Impact of fabrication errors on MBS characteristics of optimized arrays. 
The histograms compile (a) the topological gap $\Delta_0$, and (b) the MBS energy splitting $\epsilon_M$ for 300 realizations of disorder.
(c,d) Disorder-averaged values of $\Delta_0$ and $\epsilon_M$, represented by filled circles, for optimized arrays at different 2DEG depths $d$. 
The shaded areas represent the standard deviation of the distributions $\Delta_0$ and $\log_{10}(\epsilon_M/2)$.
The distribution of the zero-mode energy splitting ($\epsilon_M$) is expected to be log-normal in the limit of long wires \cite{PhysRevLett.107.196804}.}
\end{figure}

Figures~\ref{fig:optimization} (a-b) and Table \ref{tab:param gap-xcm} depict the influence of disordered arrays on the MBS parameters.
We find that certain disorder configurations lead to better MBS characteristics in geometries I and III, while all realizations of noise reduce the topological gap in geometry II. 
At any rate, the disorder-averaged gap is smaller than the optimized value of a perfect array, regardless of the geometry.
In this regard, geometry~I shows the strongest robustness against disorder, with the largest disorder-averaged gap ($\overline{\Delta_0}$), the  
lowest standard deviation for the gap ($\sigma_{\Delta_0}$) and the lowest disorder-averaged MBS energy splitting ($\overline{\epsilon_M}$). 
The relative fragility of MBS to disorder in geometry III can be understood by looking at the spatial variation of the magnetic field in Fig.~\ref{fig:3 geom} (d-f). 
Unlike in geometries I and II, small relative displacements of the magnets along the $y$ direction lead to a strong cancellation of the spiral field amplitude and account for the broad distribution of $\Delta_0$ and $\epsilon_M$. 

\begin{table}[h!]
\caption{Topological gap $\Delta_0$ and the MBS energy separation $\epsilon_M$ in perfect and disordered magnet arrays, for $d = 50$~nm. 
For perfect arrays, the optimal values ($\Delta_0^{\rm opt}$ and $\epsilon_M^{\rm opt}$) are listed.  
For disordered arrays, disorder-averaged values ($\overline{\bullet}$) and standard deviations ($\sigma_\bullet$) are shown.
\label{tab:param gap-xcm}}
\begin{ruledtabular}
\begin{tabular}{ccccccc}
Geom. \ & \ $\Delta _{0}^{\mathrm{opt}}$ \ & \ $\overline{\Delta_0}$   \ & \ $\sigma _{\Delta_0}$  \ & \ $\epsilon_{M}^{\mathrm{opt}}$ \ & \ $\overline{\epsilon_M}$ \ & \ $\sigma _{\log_{10}(\epsilon_M)}$ \ \\

 \ & \ (mK) \ & \ (mK)  \ & \ (mK) \ &  \ (nK) \ & \ (nK) \ & \ (nK) \   \\
\hline 
I \ & \ 57 \ & \  56 \ & \ 3 \ &  \ 56 \ & \ 78 \ & \ 36  \  \\
II \ & \ 59 \ & \  49 \ & \ 4 \ &  \ 580  \ & \ 2400 \ & \ 4000  \ \\
III \ & \ 70 \ & \  55 \ & \ 11 \ &  \ 1200 \ & \ 1120  \ & \ 1320  \ \\
\end{tabular}
\end{ruledtabular}
\end{table}

From the previous discussion, it appears that a tradeoff has to be made between the largest topological gap in an optimized array and its robustness to disorder. 
Depending on the precision of microfabrication protocols, design choices could be oriented either to noise resilient geometries (such as geometry I) or to geometries with the largest optimized parameters (geometry III). 
Moreover, one would need to consider how other parameters, such as the distance $d$ between the magnets and the nanowire, affect the MBS properties.

It turns out that the Zeeman energy and the effective spin-orbit coupling are greatly affected by the 2DEG depth $d$.
In shallow 2DEGs, the increased Zeeman energy and spin-orbit coupling can boost the MBS characteristics at the expense of larger magnetic field gradients that amplify the impact of disorder. 
Increasing $d$ smooths the field profiles and MBS characteristics becomes less sensitive to disorder. 
This effect is best seen in Fig.~\ref{fig:optimization}(c-d) for geometry III, where the smoother magnetic texture mitigates the impact of disorder on $\Delta_0$ and $\epsilon_M$. 
Indeed, at $d>50$ nm, geometry III presents the strongest gap and robustness to disorder. 

The reduction of $\Delta_0$ with $d$ in geometries I and II  (Fig.~\ref{fig:optimization}(c)) can be understood in terms of the gaps $\Delta_{1}$ and $\Delta_{2}$ [cf. Eqs.~\eqref{d1} and~\eqref{d2}]. 
In geometry I, $\Delta_0$  is limited by $\Delta_1$ and the lower Zeeman energy at higher $d$ further reduces the topological gap. 
In geometry II, the gap is limited by $\Delta_2$ and the strong reduction of the spin-orbit energy contributes to the low $\Delta_0$ at high $d$. 
In geometry III, the gap is roughly constant over the studied range in $d$. 

\begin{figure}[t]
\includegraphics[width=1.0\linewidth]{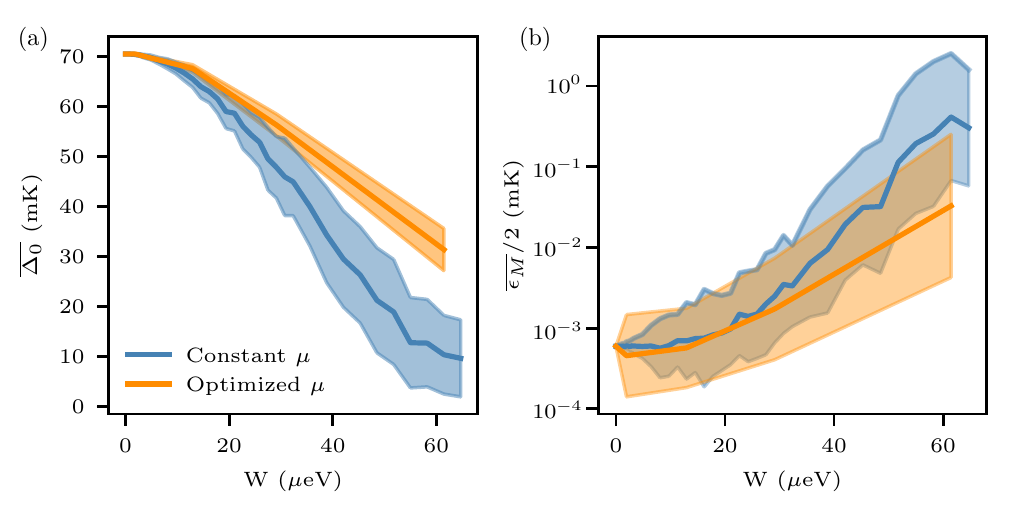}
\caption{\label{fig:charge_noise} 
Impact of random non-magnetic disorder on (a) the topological gap protecting the MBS and (b) the energy splitting of the zero energy modes, for an optimized magnet array in geometry III (see Fig.~\ref{fig:3 geom}). The blue curve displays the disorder-averaged topological gap as a function of the disorder amplitude $W$, for a fixed chemical potential. In the orange curve, the uniform part of the chemical potential is optimized for each realization of the disorder. This additional optimization cancels the shift in the chemical potential induced by disorder, thereby improving the protection of the Majorana bound states. The shaded areas represent the standard deviation of the distributions $\Delta_0$ and $\log_{10}(\epsilon_M/2)$, obtained for 100 iterations of random disorder.
}
\end{figure}
In the aforementioned discussion, we have demonstrated the robustness of MBS to experimentally relevant magnetic disorder. 
A spatially smooth disorder in the Zeeman field results in random shifts of the local energy spectrum, which (for fixed chemical potential) leads to a random and spatially inhomogeneous charge density profile.
There may in addition exist disorder of non-magnetic origin (e.g. from charged impurities located in the vicinity of the nanowire).
We model this disorder by adding to Eq.~\eqref{eq:Htot} a spatially uncorrelated random onsite potential, whose amplitude between $-W$ and $+W$ is sampled from a uniform distribution.
Figure \ref{fig:charge_noise} shows the evolution of the disorder-averaged topological gap as a function of the disorder strength $W$. 
We find that MBS remain robust up to values of $W$ of the order of the Zeeman energy scale (a few tens of $\mu{\rm eV}$). In sum, in order to have robust MBS, the uniform part of the chemical potential must be tuned with an accuracy of a few tens $\ \mu{\rm eV}$, and disorder in $\mu$ must be bounded by a few tens $\mu{\rm eV}$. The latter condition is achievable in clean wires, provided that the mean free path is greater than $\gtrsim\hbar k_f/(m^* \times  10\ \mu{\rm eV}) \sim 1-10\ \mu{\rm m}$. 
 \\


\section{Influence of material parameters on MBS attributes\label{sec:materials}}

In previous sections, we have focused our analysis on parameters that are relevant for silicon. 
In the model Hamiltonian (see Sec.~\ref{subsec:model}), the choice of material is reflected in the values of the effective mass $m^*$, the $g$ factor, and the spin-orbit coupling $\alpha$. 
The proximity-induced superconducting gap $\Delta$ is treated as a fixed phenomenological parameter.
\begin{figure}[htb]
\centering
\includegraphics[width=1.0\linewidth]{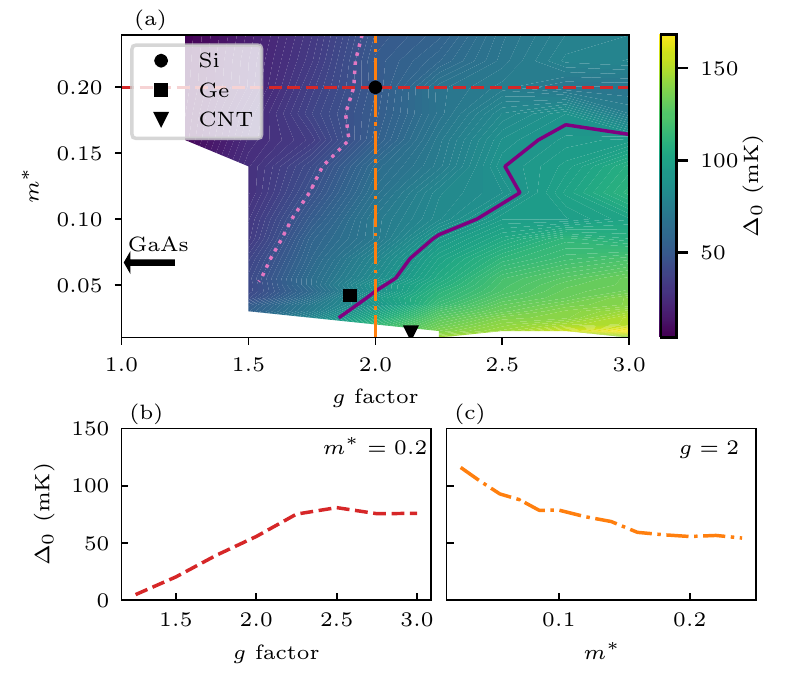}
\caption{\label{fig:meff}
(a) Topological gap $\Delta_0$ for optimized magnet arrays (geometry I, see Fig.~\ref{fig:3 geom}), as a function of the effective mass $m^*$ and the $g$ factor. 
Dotted light (solid dark) purple curve is a constant gap contour corresponding to $\Delta_0=50 (100)$~mK.
Each point of panel (a) is the outcome of an independent optimization starting from an array of cubic magnets (cf. Sec.~\ref{subsec:optim}). 
White regions identify the parameter space where the optimizer could not reach the topological phase.
Markers indicate the experimental parameter values for carbon nanotubes (CNT)~\cite{kuemmeth2008} and germanium~\cite{giorgioni2016}.
The red dashed (orange dot-dashed) curves indicate cuts for the effective mass ($g$ factor) of silicon plotted in panels (b) and (c).
}
\end{figure}

We first explore the MBS characteristics for a region of parameter space close to that of silicon, which is relevant for weakly spin-orbit coupled semiconductors ($\alpha  \approx 0$). 
Due to its experimental simplicity and robustness to nanofabrication errors (see Sec.~\ref{sec:disorder}), we focus on geometry I. 
Figure~\ref{fig:meff}(a) presents the optimized topological gap in the parameter space spanned by $m^*$ and $g$, where each point is the result of an independent optimization of the magnet array. 
Markers in Fig.~\ref{fig:meff} indicate the parameters corresponding to various materials, taken from the literature. 
In practice, one can move somewhat in parameter space by changing the material or by renormalizing parameter values e.g. through quantum confinement or through hybridization to a superconductor~\cite{Mikkelsen2018}. 

The influence of the $g$ factor on the topological gap is straightforward. 
The band inversion leading to MBS being controlled by the Zeeman energy, sufficiently large values of $g$ $(\gtrsim 1.5)$ are needed in order to drive the topological phase transition. 
However, for larger Zeeman energies, the topological gap becomes limited by the effective spin-orbit coupling strength induced by the magnet array [see line cut in Fig.~\ref{fig:meff}(b)].
As this quantity is inversely proportional to the effective mass [cf. Eq.~\eqref{eq:alpha_eff}], the largest topological gaps are found in the lower right corner of the plotted parameter space (largest $g$ and smallest $m^*$ values). This leads to constant gap contours (dotted light and solid dark purple curves) with a positive slope in the ($g$, $m^*$) space.

One caveat to the above analysis is that the effective mass also affects the superconducting coherence length $\xi$.
For a fixed $\mu$ and $\Delta$, a smaller effective mass increases the superconducting coherence length and thus the importance of finite-size effects. 
Indeed, when $E_{\rm so}\propto 1/m^*$ exceeds the Zeeman energy, the localization length of the MBS becomes proportional to $\alpha_{\rm eff}$ and thus to $1/m^*$~\cite{Aguado2017a}. 
Then, the overlap of MBS in regimes where $\Delta_0$ is small becomes particularly significant, making it difficult for the optimization procedure to reach the topological phase. This explains the negative slope of the phase transition line in the ($g$, $m^*$) space.

\begin{figure}[tb]
	\centering
	\includegraphics[width=1.0\linewidth]{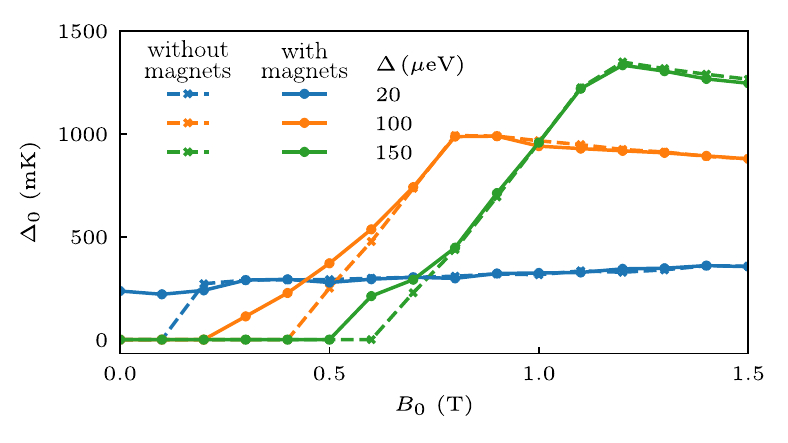}
	\caption{
	Topological gap in a single-channel InAs nanowire with (disks, solid curves) and without (crosses, dashed curves) an optimized magnet array.
	Each disk is the result of a separate optimization. We consider a magnet array in geometry I (see Fig.~\ref{fig:3 geom}) and an InAs nanowire of length $L=5$~$\mu$m, $m^*=0.023$, $g=8$, and an intrinsic spin-orbit coupling $\alpha/\hbar = 3\times 10^4$~m/s~\cite{Lutchyn:2017qy}. 
	}
	\label{fig:optim_wSOC}
\end{figure}

To conclude this section, we consider the application of magnet arrays for the engineering of MBS in materials with strong intrinsic spin-orbit coupling. 
The interplay between magnetic textures and spin-orbit coupling has been predicted to yield a complicated phase diagram that includes fractionalized fermions~\cite{Klinovaja2012b,Rainis:2014le}, as well as MBS if superconductivity is induced. 
Here, we focus on the case where the magnetic texture and the spin-orbit coupling act in cooperation. 
This is the case when the helical texture is in the plane perpendicular to the intrinsic spin-orbit interaction. 
In such a case, magnet arrays could be used to boost the effective spin-orbit coupling of the device and locally increase the Zeeman energy.

To illustrate this point, Fig.~\ref{fig:optim_wSOC} presents the topological gap for the parameters of an InAs nanowire covered by a partial shell of aluminium~\cite{Lutchyn:2017qy} and compares the case of a standard uniform wire (dashed curves) to the case where a magnet array is added to the device following geometry~I. 
For all superconducting gap amplitudes considered, the magnets allow to lower the external magnetic field amplitude needed to reach the topological phase.
This could be particularly useful in the context of hybrid nanocircuits involving, on the same chip, both Majorana-based qubits~\cite{Karzig2017,Plugge:2017dg} and superconducting circuits~\cite{Jiang:2011et,Bonderson:2011zy}. 



\section{Influence of multi-band effects}\label{sec:discussion}

In this section, we investigate the impact of multiple conduction channels in the nanowire and show that they do not significantly alter the results obtained from the one-band model.

We include multi-band effects by modifying the model described in Sec.~\ref{subsec:model} to consider the extension of the 2DEG in the $y$-direction and the confinement leading to an effective one-dimensional wire. Hence, we consider the two-dimensional Hamiltonian
\begin{equation}
	H_0  = \int_0^L \drm x \! \int_{-L_y/2}^{L_y/2}\drm y \, \psi^\dag(x,y) \Hm(x,y) \psi(x,y),
\end{equation}
where $L_y$ is the width of the effective wire. Due to the inclusion of a confinement potential $L_y\ll L$ is sufficient to capture the correct description of the low-energy sub-bands. 
The 2D extension of the non-superconducting part of the Hamiltonian is
\begin{equation}
	\Hm_0(x,y) = 
	\frac{ p_x^2 + p_y^2 }{2m^*} +V(y) - \mu + \frac{ 1 }{2}g \mu_B \Bv(x,y)\cdot \sigv , 
\end{equation}
where $V(y)$ is a confinement potential generated e.g. by electrostatic gates or etching.

In the regime of interest, where only a few sub-bands are occupied, the potential can be approximated by~\cite{Datta:1997lp}
\begin{equation}
\label{eq:conf}
V(y)=\frac{1}{2} m^* \omega_{\rm conf}^2 y^2, 
\end{equation}
which leads to an inter sub-band energy separation of $\delta E\simeq \hbar\omega_{\rm conf}$. In order for the results from the single band model to hold, the variation of the magnetic field texture in the transverse direction should be small on the scale of $l_{\rm conf} = (\hbar/m^*\omega_{\rm conf})^{1/2}$,
the electronic localization length in the transverse direction. 
This will be the case if $w$, the linear dimension of the magnets in the $y$ direction, is large compared to $l_{\rm conf}$.
Similarly, in this regime we can neglect any spatial variations of the proximity-induced superconducting gap.

As in the previous sections, we discretize $H_0$ on a square lattice leading to a $N_x \times N_y$ tight-binding model which we diagonalize numerically. Up to the small $y$-dependence of the local magnetic field, 
this is equivalent to considering $N_y$ copies of the one-dimensional wire described in Eq.~\eqref{eq:Htot} and by adding a spin-independent interwire coupling as well as a confinement potential in the transverse direction. 
In practice, $\omega_{\rm conf}$ is a complicated function of the voltage applied on the gates, the device geometry and the material parameters. We use our simulations to give an approximate lower bound on $\delta E$ below which the multi-band effects are unimportant.
We consider in our numerical calculations three values of $\omega_{\rm conf}$ between $0.05$ and $0.5$~meV, leading to $l_{\rm conf} \lesssim 90$~nm $\ll w$.

In Fig.~\ref{fig:multiband}, we have calculated the topological gap for this multi-band model, with the magnet array parameters obtained from the optimization of the single-channel model. 
Up to a shift of the chemical potential, the outcome does not change appreciably our earlier results in the case where a single sub-band is occupied.
For a modest number of occupied sub-bands, the topological gap we find is quantitatively similar to the case of a single occupied sub-band.

The robustness of the MBS under magnetic and chemical potential disorder is also similar to the case with only one occupied sub-band, provided that the inter sub-band energy separation $\delta E$ exceeds the maximal disorder amplitude. 
In the single-band limit, we have seen that the MBS that emerge in our architecture are robust under chemical potential variations of the order of $10\ \mu{\rm eV}$.
Insofar as this energy scale is small compared to $\delta E$, it follows that the results from the single-band model are not significantly affected by multi-band effects.

\begin{figure}[t]
\includegraphics[width=0.45\textwidth]{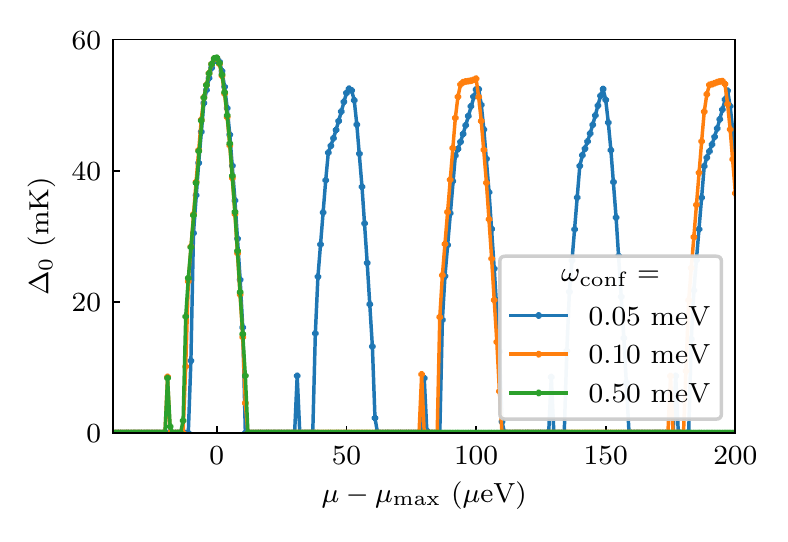}
\caption{\label{fig:multiband} 
Topological gap in a multi-channel nanowire, calculated for a width of $N_y=30$ sites ($L_y=750$~nm). The magnet arrangement corresponds to geometry I. The three colors denote different values of the transverse confinement potential (see Eq.~(\ref{eq:conf})), corresponding to localization lengths of 88~nm, 62~nm and 28~nm in the transverse direction.  The topological gap is nonzero for intervals of the chemical potential in which there is an odd number of occupied sub-bands in the normal state. The distance between subsequent topologically nontrivial regions indicates the inter sub-band energy separation. The magnitude and robustness of the optimal topological gap do not depend appreciably on the value of the confinement potential, provided that the inter sub-band energy separation exceeds the width of each topological lobe ($\sim 10\ \mu{\rm eV}$).
}
\end{figure}

For completeness, we have also rerun our optimization algorithm  using directly the multi-band model. 
Its outcome does not change appreciably our earlier results of sections \ref{sec:arrayprop}-\ref{sec:materials}.
The reason for this is that the optimization algorithm evolves the chemical potential to a value for which an odd number of sub-bands are occupied (which corresponds to the topological phase).
We have also explored the case in which we fix the chemical potential so that it intersects with multiple sub-bands. 
When only optimizing the magnetic texture, we find that the number of occupied sub-bands will remain unchanged in the course of the optimization.
The change in the energy bands due to local Zeeman fields is not sufficient to change the number of occupied sub-bands except when the chemical potential is very close to a sub-band edge. 
Indeed, the Zeeman energy corresponding to a field of $100\ {\rm mT}$ is only about $10\ \mu{\rm eV}$ (for a $g$ factor of order unity), which is small compared to the values of $\delta E$ considered in this work.\\


\section{Conclusion}\label{sec:conclusion}

In summary, our results indicate that it is possible to reach robust Majorana zero modes in low g-factor materials using magnet arrays. 
To do so, the chemical potential must be controlled with an accuracy of $\sim 10\ \mu{\rm eV}$ (which corresponds to the Zeeman energy), while its random fluctuations must be bounded by a few $\sim 10\ \mu{\rm eV}$.
Likewise, the periodic modulations of the electrostatic potential caused by the presence of the magnets must be bounded by the Zeeman energy, or else be compensated in order to ensure a sufficiently uniform electron density in the nanowire.
Materials with small effective masses appear to be promising for the realization of such Majorana devices, though, in general, the topological gaps predicted in our work are smaller than those reported in semiconducting nanowires coated with epitaxial aluminum.

An optimal Majorana device should exhibit the following main characteristics: 
(i) a large topological gap, 
(ii) well localized MBS (small wave function overlap) characterized by a small energy splitting between the zero energy modes, and 
(iii) robustness of the MBS characteristics to variations in parameters with respect to optimized nominal values.
In the case of a magnet array, there is an additional criterion: (iv) robustness to variations in the dimension and position of micromagnets that are caused by unavoidable microfabrication errors and non-magnetic disorder. 
While none of the three geometries considered in this work are ideal with regards to all of those criteria, the following assessment of their strengths and weaknesses can help guide the design of devices.

First, our results show that geometry III is optimal with respect to criteria (i and iii).
It exhibits the largest topological gap at optimized parameters, which is stable over a range of parameters around the optimized values.  
However, for 2DEGs of depth $d \leq 50$~nm, this geometry is the most sensitive to microfabrication errors and does not perform as well as geometries I and II when it comes to criterion (iv). 
Therefore, depending on the expected fabrication precision and the tolerable device yield, one might prefer geometry I where the optimal topological gap is slightly smaller, but the resilience to microfabrication errors is larger.
Second, for heterostructures in which the channel depth is over 50~nm, geometry~III appears to be most suited since its topological gap exceeds 40~mK, while for the other two geometries $\Delta_0<20$~mK. 

Overall, our work demonstrates the advantage of automated optimization as a tool for the design of Majorana devices based on magnetic textures.
Using a simple nanowire model, our results provide design guidelines for Majorana devices based on simple periodic micromagnet arrays, and allow to explore a large parameter space of magnet and material parameters. 
Our approach could be augmented to carry out a more precise optimization for specific geometries and materials by using recent developments in the advanced modeling of hybrid semiconductor-superconductor devices~\cite{Antipov2018,Winkler2018}. 


\begin{acknowledgments}
This work was funded by the Canada First Research Excellence Fund, the Natural Sciences and Engineering Research Council of Canada, and the Fonds de Recherche du Qu\'{e}bec -- Nature et Technologies. 
Numerical calculations were done with computer resources from Calcul Qu\'{e}bec and Compute Canada.
\end{acknowledgments}

\end{document}